\input{aipcheck.tex}

\documentclass{aipproc}
\layoutstyle{6x9}

\begin{document}

\title 
      [Scalar mesons]
      {Light scalar meson spectrum} 


\author{Wolfgang Ochs}{
  address={F\"ohringer Ring 6, D-80805 M\"unchen, Germany},
  email={wwo@mppmu.mpg.de},
}

\copyrightyear  {2001}

\begin{abstract}
We discuss the classification of the light scalar mesons
with mass below 2 GeV into $q \bar q $ nonets and  glueballs.
The information on
production and decay of these states, in particular recent information
on the  $f_0(980)$, $f_0(400-1200)$ (or $\sigma(600)$) 
and  $f_0(1500)$ is considered.
 Although the data are not yet very precise the recent information
is in favour of the previously developed scheme which includes 
 $f_0(980),\ a_0(980),\ K_0^*(1430),\ f_0(1500)$ 
into the lightest scalar nonet. The glueball in this approach appears as
broad object around 1 GeV. Alternative schemes find the glueball at somewhat
higher mass or suggest his mixing with $q\bar q$ states spread 
over a similar mass range. 
We do not see sufficient evidence yet for a light scalar nonet below 1 GeV
around a $\sigma(600)$ resonance.
\end{abstract}

\date{\today}

\maketitle

\section{Introduction}
Among the light  mesons  the scalars ($J^{PC}=0^{++}$) 
are quite resistant against any classification which would be generally
acceptable. 
In part this comes from the difficulty to identify very broad states, like
$f_0(400-1200)$ 
and to determine  their parameters or
even to establish their presence, 
in part also from the possible existence of different types of mesons,
 namely besides the usual $q\bar q$  quarkonia, 
four quark or molecular states and glueballs.
At the same time these possibilities are at the origin of the large interest
in the scalar states as the lightest glueball
is generally expected with these quantum numbers.
It is therefore important 
to bring order into the scalar spectrum.

The existence of glueballs is confirmed by the nonperturbative QCD
calculations and their  discovery is a challenge for theory and experiment. 
In the quenched approximation of lattice QCD, i.e. 
neglecting sea quarks, one finds the mass of the 
scalar glueball in the range 1400-1800 MeV (recent reviews
\cite{bali,morningstar}). There are still considerable uncertainties
concerning the effects of sea quarks and the masses of the light quarks.
The glueballs  with
tensor and pseudoscalar quantum numbers are next heavier in mass
and already close to or above 2 GeV.
At this higher mass it is certainly more difficult to establish the nature
of the observed resonances, therefore the scalar glueball is the primary
target for glueball searches. 
An alternative approach to glueball masses is based on QCD sum rules.
In these calculations
a light gluonic state near 1~GeV is 
demanded from a particular sum rule
\cite{narison,steele}. According to these QCD results
it seems plausible to search for the lightest scalar glueball 
in the range 1 - 2 GeV and not in a much narrower region.    

In a first step it should be clarified which states 
with $J^{PC}=0^{++}$ are really established
and what their internal flavour structure is. Then one can attempt
to group them into $q\bar q$  nonets. The existence of
glueballs is indicated if there are supernumerary states.
In the following we discuss in particular results from Ref.
 \cite{mo,mo1} including new analyses  \cite{mo2} and compare with other 
approaches.

\section{Light scalar mesons: evidence  and flavour structure}
First we attempt to identify the $q\bar q$ meson nonet(s). The scalar
states listed by the Particle Data Group (PDG) are shown 
in Table \ref{tab:scalars}. 
In this talk we concentrate our attention
to the isoscalar ($f_0$) states of lowest mass and the construction of
the lightest nonet. 

In order to establish a resonance in a general environment with background,
which case is relevant to our discussion, we request 
as necessary condition for a state to be acceptable 
a definite evidence for
the movement of the partial wave amplitude in both magnitude and phase 
according to a local Breit Wigner 
representation, i.e. 
 a pole in the complex energy plane in general above some backgound.
We begin our discussion with two well established isoscalars where the
debate concerns their intrinsic structure and continue with two others 
whose very existence we consider in doubt.

\begin{table}[b] 
\begin{tabular}{lllll}
\hline
${\bf I=0}$ &$f_0(400-1200)$ ({\rm or} $\sigma$) & $f_0(980)$\quad $f_0(1370)$ &  
     $f_0(1500)$\quad $f_0(1710)$ &  $f_0(2020) ?$\\
${\bf I=\frac{1}{2}}$ &  &                  & $K^*_0(1430)$ & $K^*(1950) ?$\\   
${\bf I=1}$ &    &  $a_0(980)$            &  $a_0(1450)$   & \\ 
\hline
\end{tabular}
\caption{Scalar mesons below 2 GeV according to Particle
Data Group \protect\cite{pdg} }
\label{tab:scalars}
\end{table}

\subsection{The $f_0(980)$ meson}
The existence of this state is well established by early phase shift
analyses~\cite{berkeley,hyams}. There is a continuing debate
on whether its internal structure corresponds to a quarkonium or rather
to a 4-quark or molecular state. We follow here the standard quark model
for simplicity as far as possible in the hope that some problems may
disappear with improved calculations. The $q\bar q$ assumption is
supported by various observations which are not natural for a complex 4q
state:
the close similarity in various production properties 
with other quarkonia 
of similar mass (like $\rho$, $\phi(1020)$, $a_0(980)$, $\eta'$, in both 
$e^+e^-$ annihilation \cite{opalf0} and $\nu p$ 
 interactions \cite{nomf0}, see recent review \cite{klempt});
the dominance of $f_0(980)$ production at large 
momentum transfer $|t|\sim 0.5$ GeV$^2$ 
in $\pi p$ collisions \cite{gamslt} which suggests a $f_0 \pi A_1$ coupling
\cite{mo}; the strong production in $D,D_s$ decays (see below) through
intermediate $d\bar d$ and $s\bar s$ states.  
This discussion will have to continue until a consistent description
of all phenomena is achieved.

In the following
we discuss the predictions from the quarkonium model
for various ratios of observables, where the dependences on the
less known intrinsic structures are expected to cancel.
These ratios depend
only on the mixing angle $\varphi_s$ in the scalar nonet which we define
through the amplitudes into strange and non-strange components
\begin{equation}
f_0(980)=\sin \varphi_s n\bar n + \cos \varphi_s s \bar s \qquad {\rm with}
\qquad   n\bar n = (u \bar u + d \bar d)/\sqrt{2}. 
\label{mixing_s}
\end{equation}
The results can ultimately be compared with corresponding predictions from 
molecular models if available. A consistent description of data in terms
of only one parameter $\varphi_s$ is then a crucial test of the 
quarkonium model.

The definition (\ref{mixing_s}) is in analogy to the common
definition for pseudoscalars
\begin{eqnarray}
\eta' &=& \sin \varphi_p n\bar n + \cos \varphi_p s \bar s  \\
\eta &=& \cos \varphi_p n\bar n - \sin \varphi_p s \bar s.  
\label{mixing_p}
\end{eqnarray}
A recent determination of the pseudoscalar mixing angle yielded \cite{fks}
\begin{equation}
\varphi_p = 39.3^\circ \pm  1.0 ^\circ.  \label{phips}
\label{phi_p}
\end{equation}
This result corresponds approximately to components
$(u\bar u, d\bar d, s\bar s)$
\begin{eqnarray}
\eta'&=\quad (1,1,2)/\sqrt{6} \quad &{\rm near~singlet}\quad (1,1,1)/\sqrt{3}
\label{mixing_etap}\\
\eta&=\quad (1,1,-1)/\sqrt{3} \quad &{\rm near~octet}\quad (1,1,-2)/\sqrt{6}
\label{mixing_eta}
\end{eqnarray}
with mixing angle $\varphi_p=35.3^\circ$ near singlet-octet angle $\varphi_p=54.7^\circ$.

We consider here three ratios of branching ratios as 
in \cite{mo} but now express them in terms of the mixing angle
$\varphi_s$. Then we can get a quantitative measure of the consistency
of the approach. We summarize here only the final results in Table
\ref{tab:ratios}.

The experimental values for the ratios $R_i$ are determined from the PDG
results where available. The ratio $R_1$ estimates the ratio
of strange and nonstrange components of the $f_0$. This ratio has 
been determined for the pseudoscalars and yielded a mixing angle
consistent with all other determinations \cite{fks}. Remarkably
the $J/\psi$-branching ratios entering 
 $R_1$ are very similar for $\eta'$ and $f_0(980)$ which
gives the first hint
towards the close similarity of the scalar and pseudoscalar multiplets.
The ratio $R_1=2$ would correspond
to the quark composition $\eta'=(1,1,2)/\sqrt{6}$ in (\ref{mixing_etap}).
Accordingly, the mixing angle $\varphi_s\sim \varphi_p$, in addition a
second solution $\varphi_{s2}$ is possible. 

The ratio $R_2$ is calculated
from the $q\bar q$ annihilation amplitudes which are proportional
to  the squares of quark charges $Q_q^2$. It is assumed here that
the $a_0(980)$ is a quarkonium as well with wave function 
$a_0=(u\bar u - d \bar d)/\sqrt{2}$. 

For the ratio $R_3$ we
assume a strange quark suppression amplitude $S=0.8\pm 0.2$ close to Ref. 
\cite{amn}. The reduced branching ratio $g_K^2/g_\pi^2$ is taken from
determinations based on measurements of both 
$K\bar K$ and $\pi\pi$ final states in central production \cite{WA102}
and with low background in large $t$ $\pi p$ collisions \cite{binnie}.

\begin{table}[t] 
\begin{tabular}{l|ccc|}
\hline
\tablehead{1}{c}{c}{}
& \tablehead{1}{c}{c}{$
    R_1={\displaystyle\frac{J/\psi\to \phi f_0}{J/\psi\to \omega f_0} }$}
&\tablehead{1}{c}{c}{$
    R_2={\displaystyle\frac{f_0\to\gamma\gamma}{a_0\to \gamma\gamma}}$}
&\tablehead{1}{c}{c}{$
    R_3={\displaystyle\frac{f_0\to K\bar K}{f_0\to \pi\pi}}$}\\
\hline
$R_{\rm exp}$ & $2.3 \pm 1.1$  & $1.3 \pm 0.6 $ & $g_K^2/g_\pi^2=2.0\pm 0.6 $ \\
$R_{\rm theor}$ 
  &{$\displaystyle \frac{p_\phi}{p_\omega} \cot^2\varphi_s$} 
  &{$\displaystyle  \frac{2}{9} \left(\frac{5}{\sqrt{2}}\sin
   \varphi_s+\cos\varphi_s\right)^2 $}
  &{$\displaystyle  \frac{2}{3} \left(\cot \varphi_s +
    \frac{S}{\sqrt{2}}\right)^2$} \\
\hline
$\varphi_{s1}$ &  $33\pm 7^\circ $ &
$25\pm 12^\circ $ & $ 41^\circ\pm 8^\circ$ \\
$\varphi_{s2}$ &  $147\pm 7^\circ $ & 
$123\pm 12^\circ $ & $ 153.8^\circ\pm 3.4^\circ$ \\
\hline
\end{tabular}
\caption{Summary of scalar mixing angle $\varphi_s$ from three
ratios of branching fractions involving the state $f_0(980)$.}
\label{tab:ratios}
\end{table}
The results for the mixing angles from the measured ratios
are nicely compatible for the small angle solution 
($\chi^2=1.4$) with
\begin{equation}
\varphi_s=35^\circ\pm 4^\circ    \label{phisol1}
\end{equation} 
whereas the large angle solution $\varphi_s=154^\circ\pm 3^\circ$ 
closer to octet is
disfavoured ($\chi^2=6.4$, Prob$\,\sim\,$1\%). 
These results are based on three ratios and have little model
dependence. Our favoured solution $\varphi_s$ is similar
to the pseudoscalar mixing angle $\varphi_p$ in (\ref{phips}) 
as already suggested in \cite{mo}. 

Our two solutions are similar to those of Ref. \cite{amn} (using
$g_K^2/g_\pi^2\sim 1.5$) whereas in Ref. \cite{aan} calculations based on
two absolute rates yielded 
the small angle solution $\varphi_s=4^\circ\pm3^\circ$
which is rejected in favour of the large angle
solution $\varphi_s=138^\circ\pm 6^\circ$.

Independent information on the relative phase of the $s\bar
s$ vs. $n\bar n$ components in the $f_0(980)$ wave function
is accessible from $D$ and $D_s$ charmed meson decays.
Consider first the decay $D_s^+\to \pi^+K^+K^-$ which shows a strong
 $\phi(1020)$ signal overlapping with 
 $f_0(980)$. The dominant decay of $D_s^+$ proceeds
through the emission of a $\pi^+$ and formation of an intermediate 
$s\bar s$ state subsequently decaying
into $\phi$ and $f_0$ states
with amplitudes $1$ and $\cos\varphi_s$ according to (\ref{mixing_s}).
The absolute phases of $\phi$ and $f_0$ have been determined by the E687
Collaboration \cite{E687b} as $(178\pm20\pm24)^\circ$
and $ (159\pm 22 \pm 16 )^\circ$, i.e. the relative phase is consistent
with zero degrees. Therefore 
\begin{equation}
\cos\varphi_s>0 \qquad \to \qquad 0<\varphi_s<90^\circ  \label{phif0rp}    
\end{equation}
 in agreement with the small phase solution (\ref{phisol1}).

The determination of the relative phases depends  on the definition
of the scattering angle. Under an exchange of the two particles of the decay
the S-P
wave interference term considered here would change sign. As a check we
therefore studied a similar situation in the decay $D^+\to \pi^+\pi^-\pi^+$
with the relative phase between the amplitudes
$D^+\to \pi^+\rho^0$ and $D^+\to \pi^+f_0(980)$. According to the dominant
mechanism these
resonances are produced through intermediate $d\bar d$ states with
amplitudes $-1/\sqrt{2}<0$ and $\sin \varphi_s /\sqrt{2}>0$ respectively,
so one expects a $180^\circ$ phase difference. This expectation is in fact 
verified by
the measurements by both E687 \cite{E687b}  and 
E791 Collaborations \cite{E791,cg} which confirms that
the standard definition of the angles yields consistent results.

Another interesting ratio is $R=(\phi\to a_0(980)\gamma)/(\phi\to
f_0(980)\gamma)$. 
As the decays involve two decay mechanisms with or without $s\bar s$
annihilation we have no straightforward prediction. An 
explanation is possible in terms of $a_0-f_0$ mixing \cite{ck}.

\subsection{The $f_0(1500)$ meson}

This state can be considered as well established by now also. In 
$p\bar p \to 3\pi$  the Dalitz plot has been fitted with some phase sensitivity
and the S wave nature has been demonstrated (CBAR Collaboration \cite{cbar}).
Meanwhile various branching ratios became known.

The phase movement has also been seen in the Argand diagrams of
$\pi\pi\to K\bar K$ and $\pi\pi\to \eta \eta$ as obtained from the $\pi N$
production experiments which have been reconstructed using the data on
$|S|,\ |D|$ and $\phi_{SD}$ together with Breit Wigner fits to the tensor
mesons which provide the absolute phase \cite{mo}. 
The comparison of both reactions
has also demonstrated through its interference with the tensor
mesons that
\begin{equation}
T(\pi\pi\to f_0(1500) \to K\bar K) = -T(\pi\pi\to f_0(1500) \to \eta \eta)
\label{signf0}
\end{equation}
which implies that the $f_0(1500)$ has an opposite sign of the
$n\bar n$ and $s\bar s$ components \cite{mo}. A similar Argand diagram has
been obtained for $\pi\pi\to K\bar K$ \cite{cohen}; the opposite orientation of
both amplitudes in (\ref{signf0}) is also visible in the energy dependent
fits in \cite{anis00} although with different overall phase.
Unfortunately, the elastic 
$\pi\pi$ scattering is not yet uniquely determined in this region.

Further interesting information on this meson can again be obtained from decays of
$D$ and $D_s$ charmed mesons. In the decay 
$D_s\to \pi\pi \pi$ the scalars
$f_0$ can be produced through the intermediate process $s\bar s \to \pi\pi$.
This favours intermediate states with large $s\bar s$ - $n\bar n$ mixing.
One observes a strong signal from $f_0(980)$ which proves again its strong
$s\bar s$ component
 and a higher mass state
related to $f_0(1500)$ by E687 \cite{E687b} and to $f_0(1370)$
by E791 \cite{E791}. The signal is strongest near the edge of phase space
in the Dalitz plot where the two resonance bands cross 
but the mass and width appear to be closer
to $f_0(1500)$. Ultimately the study of branching ratios of this state 
has to decide.
For the time being we take this state as $f_0(1500)$.

An interesting feature common to both experiments is the large
relative phase consistent with $180^\circ$ between the production
amplitudes of $f_0(980)$ and $``f_0(1500)"$ which we interpret as
\begin{equation}
T(s\bar s\to f_0(980)) = -T(s\bar s \to f_0(1500)).
\label{signssf0}
\end{equation}
This negative phase in the fit to the Dalitz plot obviously 
corresponds  to a lack of
enhancement at the off diagonal crossing point
of the two resonance bands in this plot
which contrasts the strong enhancement
in the diagonal crossing points.

The mass of $f_0(1500)$ is close to the glueball mass obtained
in quenched approximation of lattice QCD. This at first has lead to models 
with close connection between these two states
\cite{ac}; further studies now prefer mixing models
where the superposition of the glueball and nearby $q\bar q$ mesons
correspond to the physical states 
 $f_0(1370),\ f_0(1500)$ and $f_0(1710)$ (for overview, see
\cite{klempt}). As an example we quote a recent result motivated by lattice 
calculations on mixing \cite{lw,morningstar}
\begin{equation}
f_0(1500)\ =\ -0.36\, n\bar n +0.91\, s\bar s -0.22\, {\rm glueball}.
\label{gbmix}
\end{equation}
In this example the glueball component of the $f_0(1500)$
 has a weight of only $\sim$5\%. 

Contrary to a single glueball which would mix with the flavour singlet
we see in (\ref{gbmix}) $n\bar n$ and $s \bar s$ with opposite sign.
This octet type flavour mixing is in line with our findings 
(\ref{signf0}),(\ref{signssf0})
and appears as $``$robust result$"$ in fits of the above kind \cite{close}.
On the other hand, our finding (\ref{signf0}) 
 not only requires an octet type
flavour mixing but also provides an upper limit to the
glueball contribution; 
this contribution  would add with 
the same sign to all pairs of pseudoscalars. This is an important additional
limitation to such fits not yet taken into account so far.

If we take these observations together, especially the large components 
of both
$n\bar n$ and $s\bar s$ (suggested from $D_s$ decays) and their negative
relative sign, then $f_0(980)$ and $f_0(1500)$ look like the orthogonal
isoscalar members of the $q\bar q$ nonet. We will come back to this idea below. 

\subsection{The $f_0(400-1200)$ and the $\sigma(600)$ meson}
This entry in the PDG refers to results from $\pi\pi$ phase shift analyses
and from the observation of peaks in mass spectra around 400-600 MeV.
We give a short account of these observations.

\subsubsection{$\pi\pi$ phase shifts and $f_0(400-1200)$}
It is a common feature of fits to $\pi\pi$ scattering that there
is one broad object where the width is comparable to the mass 
(see \cite{mo}, for example).
The  $\pi\pi$ scattering amplitudes 
are rather well known by now
up to $\sim$ 1.4 GeV, from single pion production with and without
polarized target. Recent studies
have removed remaining ambiguities \cite{klr} 
below 1~GeV in favour of a slowly rising
S wave phase shift in the $\rho$ region, excluding in particular a 
rapidly varying phase and resonance under the $\rho$,
in essential agreement with the old results which were obtained using
particular assumptions on the production mechanism \cite{hyams}. 
A theoretical analysis \cite{acgl} based on the 
constraints from S matrix theory provides a good description
of the observed low energy (below 1~GeV) pion-pion interactions
with slowly varying S wave.

The interpretation of the $I=0$ S wave
in terms of resonant states is less straightforward.
The phase shifts pass through $90^\circ$ at $\sim$1000 MeV once the
$f_0(980)$ effects are subtracted. This suggests a state at 1000 MeV
\cite{mp,mo} with a large width of 500-1000 MeV. With a negative background
phase added the resonance position can be shifted towards lower values
and this has been considered as state $\sigma(600)$ in \cite{iii}. Fits
over a large mass region including a background term yield resonance poles
in the scattering amplitude around 1300 MeV or higher, again with a large
width \cite{anisres}. With such broad states the determination of the 
resonance mass
depends on the assumed background in an essential way. There is a strong
$\pi\pi$ interaction around 1 GeV and beyond but not necessarily and
exclusively a broad $\sigma(600)$. 

\subsubsection{Peaks in mass spectra and $\sigma(600)$}
There are a number of effects which have been related to $\sigma(600)$.

1. Decay $J/\psi\to \omega\pi\pi$\\
There is a peak around 500 MeV in the $\pi\pi$ mass spectra
besides a strong signal from $f_2(1270)$
\cite{dm2}.
For a Breit Wigner $\sigma$ resonance at 500 MeV 
the interference term $Re(SD^*)$
between the (almost real) D wave and the resonant S wave would change sign
at the pole position and so 
the angular distribution
$d\sigma/d\Omega
\sim |S|^2 + (3\cos^2\vartheta-1) Re(SD^*)/2 + \mathcal{O}(|D|^2)$ 
would vary
accordingly with a sign change of the $\cos^2\vartheta$ term (from + to --).
The data \cite{dm2}
do not show any sign change below 750 MeV and therefore there is
no indication for a Breit Wigner resonance below this mass.

2. $Y',Y'' \to Y\pi\pi$ and similar decays of $J/\psi$\\
Mass peaks are observed here as well. Unfortunately the angular
distributions are not measured as function of the mass.
Hopefully such measurement will be provided in the future to study
possible phase variations and resonant behaviour.

3. Central production $pp\to p(\pi\pi)p $\\
At small momentum transfers between the protons this process is 
assumed to be dominated by double Pomeron exchange.
The centrally produced $\pi\pi$ system peaks shortly above threshold
below 400 MeV \cite{afs,gamspipi,wa102pp,gams} and has been related to the
$\sigma(600)$ as well \cite{gamspipi}. There are some other remarkable
features in this process. Quite unusually, there is a strong D wave
near threshold as well which peaks near 500 MeV; the total D 
wave contributions $\sum_\lambda |D_\lambda|^2$ at their peak are about
five times larger than the $f_2(1270)$ contribution  and 
about one third of the 
S wave contribution at its peak.

These observations are very similar to findings in $\gamma\gamma\to \pi \pi$
which suggest a close relation between the processes \cite{mo2}
\begin{equation}
{\rm Pomeron~Pomeron} \to \pi \pi\qquad  \leftrightarrow \qquad
             \gamma\gamma \to \pi \pi
   \label{pomgam}
\end{equation}
In fact, the $I=0$ S wave component obtained from a fit to 
$\gamma\gamma\to \pi \pi$ for charged and neutral pions \cite{bp}
peaks below 400 MeV and the D wave near 500 MeV with similar ratio 1/3;
 the origin of this
unusual behaviour is the contribution of one-pion-exchange to 
$\gamma\gamma\to \pi^+ \pi^-$.
Therefore we propose that one-pion-exchange dominates
the double Pomeron reaction at small $\pi \pi$ masses as well
\cite{mo2}.\footnote{%
Some global properties of this process have been considered already
 long ago \cite{vak}.}
It reproduces the main characteristics.
As the pion pole is near the physical
region the  $\pi \pi$ angular distribution is very steep, steeper than
in more typical
interactions mediated by vector ($\rho$) exchange. Therefore 
one estimates that the D wave
becomes important not at $m_{f_2}$ but already 
at $m_{f_2} \times (m_\pi/m_\rho) \sim 0.3$ GeV. This mechanism also
explains the low mass peak of the S wave without associated phase variation.
In fact, in the region of the peak no strong S-D phase variation   
is observed \cite{wa102pp,gams} as would be expected for a $\sigma(600)$
resonance.
The presence of the one-pion-exchange process does not exclude
the presence of broad states as in $\pi \pi$ elastic scattering  either from
$\pi \pi$ rescattering or by direct formation, very much as
it is discussed in the $\gamma\gamma$ process \cite{bp}; but there is no
evidence for an additional low mass $\sigma(600)$ resonance near the peak. 

4. Decay $D^+\to \pi^-\pi^+\pi^+$\\
The $\pi^+\pi^-$ mass spectrum presented by the
E791 Collaboration \cite{E791,cg} shows three prominent peaks,
one just above $\pi\pi$ threshold, one related to  $\rho$ and one to
 $f_0(980)$.
Only
fits including a light $\sigma$ particle have been found successful according
to their analysis. In principle, the low mass peak could 
be due to the decays of  resonances with higher spin
$J\geq 1$ in the crossed channel. 

Again, one would like to see 
the related Breit Wigner phase motion in a more direct way. There should
be a large term $Re(SP^*)$ from the interference 
 $\sigma-\rho$ which changes sign near the $\sigma$ resonance in 
$s_{12}(\pi^-\pi^+_1)$ and
therefore
changes sign of the forward backward asymmetry in the $\sigma$ decay angle
$\cos\vartheta$ which is linearly related to the mass variable  
$s_{13}(\pi^-\pi^+_2)$
along the $\sigma$ resonance $(s_{12})$ band in the Dalitz plot. Such an effect
is actually visible in the Dalitz plot at the $\rho$ resonance
(presumably from the ($\rho-f_0(980)$ interference): the sign change
of the asymmetry causes the appearence of 
the tails towards lower and higher mass ($s_{12}$) 
 at the upper and lower part ($s_{13}$) of the 
$\rho$ band respectively. 
A study with sufficiently fine binning should 
reveal this effect for the $\sigma$
in the resonance Monte Carlo for the $\sigma+\rho$ 
superposition and prove or
disprove the presence of the interference effect in the data (a sensitive
observable is $\langle\cos\vartheta\rangle\frac{d\sigma}{ds_{12}}$
together with  $\frac{d\sigma}{ds_{12}}$).

\subsection{The $f_0(1370)$ meson}
There is strong interaction in $\pi\pi$ and other channels 
in this mass region but again the clear evidence for a 
localized Breit Wigner phase
motion is missing to support
the resonance hypothesis.
The $f_0(1370)$ and $f_0(400-1200)$ look like
parts of a broader state in the channels with two pseudoscalars
\cite{mo}. The missing information could be
provided by phase shift analysis of recent high statistics
$\pi^0\pi^0$ data \cite{gamslt,bnlpipi}.
Furthermore,
the resonance interpretation is found not consistent \cite{klempt} 
with the different
branching ratios observed in the $4\pi$ channel;
there could be a broad background state interfering
with the narrow $f_0(1500)$ to cause a peak near 1370
MeV. Again a phase shift analysis is necessary to clarify the situation.

\section{Construction of the lightest quark anti-quark nonet} 

For the time being neither the  $\sigma(600)$ nor the $f_0(1370)$ are
acceptable for us as genuine resonant states. There are 
peaks at these masses but the associated Breit Wigner phase variation 
has not yet been established.
Then 
the most natural candidates for the lightest $q\bar q$ nonet are 
\begin{equation}
f_0(980),\quad  f_0(1500), \quad K^*_0(1430), \quad a_0(980)
\label{nonet}
\end{equation}
with mixing pattern very similar to the one observed in the pseudoscalar nonet 
\begin{equation}
f_0(980)\sim \eta'\sim {\rm singlet}\qquad 
 f_0(1500)\sim \eta\sim {\rm octet}.
\label{eta_f0}
\end{equation}
A solution like this has been proposed on the basis of a 
renormalizable sigma model with
instanton interactions \cite{instanton}. It also has $f_0(1500)$ as octet
member 
but would prefer $a_0(1450)$ over $a_0(980)$
as isovector. It explains why the octet state
is above the singlet for the scalars and vice versa for the pseudoscalars.
Similar models have been studied in \cite{instanton2,bgo}.
 
Independently, the correspondence (\ref{eta_f0})
and the nonet (\ref{nonet}) have been proposed 
on the basis of the phenomenological analysis \cite{mo}. These results
were found consistent with 
a general QCD potential model for sigma variables; in this
analysis the choice $m(a_0(980))\approx m(f_0(980))$ is possible
although not required or explained. Furthermore, it has been shown that the
octet in (\ref{nonet}),(\ref{eta_f0}) fulfills approximately the Gell Mann Okubo
formula, and from $a_0$ and $K^*_0$ one predicts 
\begin{equation}
3(m_{f_8}^2-m_a^2) = 4(m_{K^*_0}^2-m_a^2)\qquad \to \qquad m_{f_8}=1550\ 
{\rm MeV} \label{gmo}
\end{equation}
a good result for the octet isoscalar.

In the low mass region we are now left with $f_0(400-1200)$ and $f_0(1370)$
to which we come back below. It is interesting to note that the remaining states
in the PDG below 2 GeV (Table \ref{tab:scalars}) can be grouped together
into a second nonet which includes
\begin{equation}
f_0(1720),\quad  f_0(2020), \quad K^*_0(1950), \quad a_0(1450).
\label{nonet2}
\end{equation}
In this case the Gell Mann Okubo formula predicts for the octet
scalar $m_{f_8}= 2.080$ GeV  which fits to the highest state in
(\ref{nonet2}) and therefore this nonet 
would repeat the mixing pattern
of the lowest nonet. However as there is little further information on these
states this assignment is rather speculative. 

\section{Candidate for lightest glueball}

After having selected the lightest nonet from well established resonances 
we are left with
$f_0(400-1200)$ and $f_0(1370)$. 
The $\pi\pi$ data are consistent with the view that both states correspond
to the low and high mass tails of a single resonance 
(``red dragon$"$ \cite{mo}) and this state we take
as the glueball
\begin{equation}
f_0(400-1200)\quad {\rm and}\quad f_0(1370) \qquad \to \qquad gb(1000).
\label{glueball}
\end{equation}
This mass corresponds to a resonance fit of $\pi\pi$ 
elastic scattering without
background, other fits have lead to higher masses with the option of a broad
glueball near 1400 MeV \cite{anisres}.
These
results are a bit lower than expected from the lattice results in quenched
approximation but looking
at the large width and the still approximate nature of 
QCD results there is not necessarily a contradiction. 

Our detailed arguments in favour of this glueball assignment have been
summarized elsewhere \cite{mo1,mo2}, together with plausible arguments for 
a large width of an S wave binary glueball.
Here we only recall the most relevant observations.\\
1. The state $gb(1000)$ is produced in most reactions which are
considered as gluon rich:\\
a. central production $pp\to pXp$;\\
b. Decays of radially excited heavy quarkonia like 
   $Y',Y'' \to Y(\pi\pi)$;\\
c. $p\bar p\to 3\pi$\\
d. There is no prominent signal however in $J/\psi\to \gamma\pi\pi$
for $m_{\pi\pi}< 1$  GeV,
this could possibly be due to instrumental problems
at small masses.\\
2. The production in $\gamma\gamma$ collisions
is untypically small \cite{mo1} (based on fits \cite{bp}).\\
3. The decays of $f_0(1370)$ (part of glueball) favour the $gb$ over the 
$n\bar n$  assignment \cite{mo,seth}.

Alternatively one may attempt to explain the 
strong $\pi\pi$  interaction in the 1~GeV
region without direct channel resonance in terms of $\rho-f$ exchange
processes ($\rho$ alone would not explain the strong $\pi^0\pi^0$
interaction) and to obtain the moving phase from
a unitarization procedure (see review \cite{klempt}). 
In considering this proposal we note that
a t-channel analysis of $\pi\pi$ scattering \cite{quigg} indicates a large
component
in the $I_t=0$ channel which is not related to $q\bar q$ Reggeon exchange but
indicates Pomeron exchange or gluonic
interactions already for $m_{\pi\pi} < 1$ GeV \cite{mo1,mo2}.
So in this explanation one has to take into account the presence of non-resonant
effects and  avoid double counting of direct and crossed channel exchanges
where the fit has to be done to all isospin amplitudes.

\section{Summary and Conclusions}
We have presented the arguments to include the 
isoscalars $f_0(980)$ and $f_0(1500)$
in the lightest scalar nonet.
Various ratios of
branching fractions as well as new results on relative phases
between different $q\bar q$ components and between
 the production amplitudes 
could be explained consistently in terms of one parameter,
the scalar mixing angle $\varphi_s=35^\circ\pm 4^\circ$. This is found close
to the pseudoscalar equivalent but very different from ``ideal mixing$"$
with $\varphi_s=0^\circ,180^\circ$. The nonet (\ref{nonet}) fulfills the Gell
Mann Okubo formula which relates the masses of octet particles
 assuming symmetry
breaking by quark mass terms.
It will be important to improve the accuracy of these measurements as test
of this classification scheme. Also it will be interesting to see whether
the $K\bar K$/ 4q model for $f_0/a_0(980)$ can explain the data discussed
here.

We do not see evidence yet for the Breit-Wigner 
resonance nature of peaks related
to $\sigma(600)$. It will be important to investigate the phase motion in
$D\to``\sigma" \pi$.
In some cases there is counter
evidence for the expected phase motion 
(like $J/\psi\to \omega\pi\pi$ and $pp\to p(\pi\pi)p $). 
Therefore the existence of a meson multiplet below 1 GeV is not
apparent.
Also there is a lack of evidence for $f_0(1370)$ so far.
There are data available which could be analysed in this respect
(for example $\pi^- p \to \pi^0\pi^0 n$). 

The PDG listing allows for a second scalar nonet below 2 GeV 
with similar mixing.

This leaves the broad state around 1 GeV (built from $f_0(400-1200)$ and
$f_0(1370)$ + more (?)) with the large width of 500 - 1000 MeV 
as a candidate for the lightest scalar glueball. This assignment is 
in agreement with most phenomenological expectations.
Also we do not see an obvious  disagreement with QCD results taking
into account the approximate nature of the calculations.

In alternative approaches a superposition of  glueball and two neighbour
scalar quarkonia (from nonet) builds up  the physical states
$f_0(1370)$, $f_0(1500)$ and $f_0(1710)$. In effect this would imply
that the glueball is not localized in a narrow mass interval but 
contributes to scattering
processes in quite a large mass range of $\sim$500 MeV as 
it happens in our ``broad glueball$"$ scheme.
Major differences are in the $q\bar q$ sector,
especially concerning the existence of $f_0(1370)$ and treating
the left alone states $f_0,a_0(980)$ as molecules or forming an additional
nonet around $``\sigma(600)"$.  Therefore improved
experimental data and analyses in the low energy region are of great
importance.

Of course it would be nice to have a more direct
evidence for the gluonic nature of a particular candidate. A promising tool
is the comparative study of glueball candidates
in the fragmentation region of both quark and gluon jets \cite{mo4,rs}. 
First results \cite{mandl} look promising in indicating an extra neutral
component in the gluon jet. There is something to look forward to.

\begin{theacknowledgments}
I would like
to thank Peter Minkowski for discussions and collaboration on the
problems presented in this talk.
\end{theacknowledgments}


\begin{thebibliography}{99}
\bibitem{pdg} 
Particle Data Group, D.E. Groom {et al.}, {\it Eur. Phys. J. C}{\bf 15}, 
    1 (2000).

\bibitem{bali}
G.S. Bali, ``Glueballs: Results and Perspectives from the Lattice$"$,
hep-ph/0110254.

\bibitem{morningstar}
C.J. Morningstar, this conference.

\bibitem{narison}
S. Narison, {\it Nucl. Phys. B} {\bf 509} (1998) 312.

\bibitem{steele}
T.G. Steele and D. Harnett, ``Two Topics in QCD Sum-Rules$"$,
hep-ph/0108232.

\bibitem{mo}
P. Minkowski and W. Ochs, {\it Eur. Phys. J. C} {\bf 9}, 283 (1999). 

\bibitem{mo1}
P. Minkowski and W. Ochs, {\it Proc. Workshop on Hadron Spectroscopy},
Frascati, March 1999, Italy,  Eds. T. Bressani et al., Frascati Physics
Series XV, p.245 (1999).

\bibitem{mo2}   
P. Minkowski and W. Ochs, in preparation.

\bibitem{berkeley}
S.D. Protopopescu {et al.}, {\it Phys. Rev. D} {\bf 7}, 1279 (1973).

\bibitem{hyams}
 B. Hyams et al., {\it Nucl. Phys. B} {\bf 64}, 4 (1973);
 W. Ochs, 
thesis 1973 (unpublished).

\bibitem{opalf0}
OPAL Collaboration, K. Ackerstaff et al., {\it Eur. Phys. J. C} {\bf 4}, 19 
    (1998). 

\bibitem{nomf0}
NOMAD Collaboration, P. Astier {et al.}, {\it Nucl. Phys. B} {\bf601},
3 (2001). 

\bibitem{klempt}
E. Klempt, ``{\it 
      Meson Spectroscopy: Glueballs, Hybrids and $Q\bar Q$ Mesons}$"$,
hep-ex/0101031.

\bibitem{gamslt} 
GAMS Collaboration: D. Alde et al. {\it Z. Phys. C} {\bf 66}, 375 (1995);
{\it Phys. Atom. Nucl.} {\bf62} 1993 (1999).
\bibitem{fks}
T. Feldmann, P. Kroll and B. Stech, {\it Phys. Rev. D} {\bf58} 114006
(1998). 

\bibitem{amn}                                                   
V.V. Anisovich, L. Montanet and V.N. Nikonov, {\it Phys. Lett. B} {\bf480}
19 (2000).

\bibitem{WA102}
WA102 Collaboration: D. Barberis  et al.,
{\it Phys. Lett. B} {\bf462}, 462 (1999). 

\bibitem{binnie}
D.M. Binnie et al., {\it Phys. Rev. Lett} {\bf 31}, 1534 (1973).


\bibitem{aan}
A.V. Anisovich, V.V. Anisovich and V.A. Nikonov,
hep-ph/0011191. 

\bibitem{E687b}
E687 Collaboration, P.L. Frabetti {et al.},  
{\it Phys. Lett. B} {\bf351}, 591 (1995); 
%
{\bf407}, 79 (1997).  


\bibitem{E791}
E791 Collaboration, 
E.M. Aitala et al., 
 {\it  Phys. Rev. Lett.} {\bf 86}, 770 (2001);
%
{\bf86}, 765 (2001). 

\bibitem{cg}
Carla G\"obel, this conference; hep-ex/0110052.

\bibitem{ck}
F.E. Close and A. Kirk,
{\it Phys. Lett. B} {\bf489}, 24 (2000). 

\bibitem{cbar}
Crystal Barrel Collaboration: V.V. Anisovich {et al.}, 
{\it Phys. Lett. B} {\bf323}, 233 (1994). 

\bibitem{cohen} 
D. Cohen et al.,  {\it Phys. Rev.} {\bf D22}, 2595 (1980).

\bibitem{anis00}
V.V. Anisovich, Yu.D. Prokoshkin  and A.V. Sarantsev,
      {\it Phys. Lett.} {\bf B389}, 388 (1996).

\bibitem{ac}
C. Amsler and F.E. Close, {\it Phys. Rev.} {\bf D53}, 295 (1996);
                           {\it Phys. Lett.} {\bf B353}, 385 (1995).

\bibitem{lw}   
W. Lee and D. Weingarten,  {\it Phys. Rev. D} {\bf 61}, 014015 (2000).

\bibitem{close}
F.E. Close, {\it Acta Phys. Polon. B} {\bf 31}, 2557 (2000). 

\bibitem{klr}
R. Kami\'nski, L.Le\'sniak and K. Rybicki, 
hep-ph/0109268.

\bibitem{acgl}
B. Ananthanarayan, G. Colangelo, J. Gasser and H. Leutwyler,
   {\it Phys. Rep.} {\bf353}, 207 (2001).

\bibitem{mp}  
 D. Morgan and M. R. Pennington, {\it Phys. Rev.}
{\bf D48} (1993) 1185.

\bibitem{iii}
S. Ishida {et al.}, {\it Prog. Theor. Phys.} {\bf 95}, 745 (1996).

\bibitem{anisres}
V.V. Anisovich et al., 
  {\it Phys. Atom. Nucl.} {\bf 60}, 1410 (2000). 

\bibitem{dm2}
DM2 Collaboration, J.E. Augustin {et al.}, 
   {\it Nucl. Phys. B} {\bf320}, 1 (1989).

\bibitem{afs}
Axial Field Spectrometer Collaboration, T. Akesson {et al.}, 
{\it Nucl. Phys. B} {\bf264}, 154 (1986).

\bibitem{gamspipi}
GAMS Collaboration,
D. Alde {et al.}, {\it  Phys. Lett. B} {\bf 397}, 350 (1997).

\bibitem{wa102pp} 
WA102 Collaboration, D. Barberis et al. {\it Phys. Lett. B} {\bf453},
325 (1999); {\bf 453}, 316 (1999). 
\bibitem{gams}

GAMS Collaboration, R. Bellazzini {et al.},
  {\it Phys. Lett. B} {\bf467}, 296 (1999).

\bibitem{bp}
M. Boglione and M.R. Pennington, 
{\it Eur. Phys. J. C} {\bf 9}, 11 (1999).


\bibitem{vak}
Ya.I. Azimov, E.M. Levin, M.G. Ryskin and V.A. Khoze, {\it Yad. Fiz.} {\bf
21}, 413 (1975). 


\bibitem{bnlpipi}
E852 Collaboration: J. Gunter {et al.},
{\it Phys. Rev. D} {\bf 64}, 072003 (2001). 

\bibitem{instanton}
E. Klempt, B.C. Metsch, C.R. M$\ddot {\rm u}$nz and H.R. Petry,
{\it  Phys. Lett. B} {\bf 361}, 160 (1995).

\bibitem{instanton2}
V. Dmitra$\check {\rm s}$inovi$\acute {\rm c}$, {\it Phys. Rev. C} {\bf 53},
 1383 (1996).

\bibitem{bgo}
L. Burakovsky and T. Goldmann, {\it Nucl. Phys. A} {\bf 628}, 87 (1998).

\bibitem{seth}
K.K. Seth, {\it Nucl. Phys. B (Proc. Suppl.)} {\bf 96}, 205 (2001).

\bibitem{quigg}
C. Quigg, in: {\it Proc. 4th Int. Conf. on Experimental Meson Spectroscopy},
Boston, 1974 (AIP Conf. Proc. no. 21, particles and fields subseries no. 8)
p. 297.

\bibitem{mo4}
P. Minkowski and W. Ochs, {\it  Phys. Lett. B} {\bf 485}, 139 (2000). 

\bibitem{rs}
P. Roy and K. Sridhar, JHEP 9907, 013 (1999).

\bibitem{mandl}
B. Buschbeck and F. Mandl (DELPHI Collaboration), 
``Study of Gluon Fragmentation and Color Neutralization$"$,
Intern. Symposium of Multiparticle Dynamics Sept. 2001, Datong, China.
\end{thebibliography}
\end{document}